\documentclass{statsoc}
\usepackage{natbib,graphicx,amsfonts}

\title[Generating Multivariate Poisson Data]{An Elegant Method for Generating \\ Multivariate Poisson Random Variables}

\author[Yahav and Shmueli]{Inbal Yahav}
\coaddress{Inbal Yahav, Department of Decision and Information Technologies, Robert H. Smith School
of Business, University of Maryland, College Park, MD 20742}
\address{Department of Decision and Information Technologies, Robert H. Smith School of Business, University of Maryland, College Park, MD
20742.} \email{iyahav@rhsmith.umd.edu}
\author[Yahav and Shmueli]{Galit Shmueli}
\address{Department of Decision and Information Technologies, Robert H. Smith School of Business, University of Maryland, College Park,
MD 20742.}

\begin{document}

\begin{abstract}

Generating multivariate Poisson data is essential in many applications. Current simulation methods
suffer from limitations ranging from computational complexity to restrictions on the structure of
the correlation matrix. We propose a computationally efficient and conceptually appealing method
for generating multivariate Poisson data. The method is based on simulating multivariate Normal
data and converting them to achieve a specific correlation matrix and Poisson rate vector. This
allows for generating data that have positive or negative correlations as well as different rates.
\end{abstract}


\section{Introduction}

The most common multivariate distribution in the statistics
literature is the multivariate Normal (\emph{Gaussian})
distribution. Generating multivariate Normal data is relatively easy
and fast. It has been used for many purposes and in a vast number of
applications. In many applications, however, the multivariate data
that arise are not well approximated by a multivariate Normal
distribution. Low count data are one example. Highly skewed data are
another. In such cases the process generating the data is better
described by a Poisson process and the resulting data by a
multivariate Poisson distribution. Such data arise in many
disciplines: In healthcare, for example, multivariate Poisson data
arise when looking at multiple series of patient arrivals, each
related to a different disease (or at different hospitals within the
same geographic area). In marketing and management science (e.g.,
pricing, revenue management) the multivariate Poisson distribution
can be used to simulate purchases of complementary products or of
substitutes. In the latter case, the correlation between purchases
of substitute products is expected to be negative.

Apart from the above examples, simulating multivariate Poisson data
is also useful for different functions such as fitting distributions
to multivariate data (fitting to a multivariate Poisson, see e.g.,
\cite{Karlis03}) and evaluating performance of statistical models by
creating simulated data. Previous studies have proposed methods for
generating bivariate and multivariate Poisson vectors based on the
joint probability function. However, such methods tend to be too
complicated for actual implementation and  are therefore rarely used
for purposes of simulation. The popular alternative is to use a
multivariate Normal approximation, which is not always adequate.

In this paper we present a new method for generating multivariate
Poisson vectors based on multivariate Normal distribution. The
method is elegant in terms of its conceptual simplicity and its
computational efficiency. Yet, it is powerful enough to allow a
flexible correlation structure, including negative and positive
correlations and unequal rates. Our method can be considered a
generalization of the univariate case by incorporating a correlation
structure that is based on a multivariate Normal formulation.

The paper is organized as follows: In Section \ref{sec:normal} we
describe the classic approach for generating multivariate Normal
data. In Section \ref{sec:poisson} we survey existing methods for
generating multivariate Poisson data and then introduce our new
method. Section \ref{sec:results} provides a graphical illustration
of the Poisson data generated by our method and their properties. We
conclude in Section \ref{sec:conclusion}

\section{Generating Multivariate Normal Data}\label{sec:normal}

We consider the common $p$-variate Normal distribution with mean
$\vec{\mu}$ and covariance matrix $\Sigma$ (denoted
$N(\vec{\mu},\Sigma)$). Simulating approaches for multivariate
Normal vector have been first addressed by \cite{Ripley87}.

\cite{lgyi98} introduced the following algorithm for generating a
$p$-dimensional Normal vector, based on the Central Limit Theorem:

\begin{enumerate}
  \item Generate $\vec{X}$, such that $\vec{X}=12\vec{U}(0,1)-6$, where
  $\vec{U}(0,1)$ denotes a vector of uniform (0,1) variates. According
  to the Central Limit Theorem $\vec{X}$ is approximately $N(\vec{0},I)$.
  \item Let $\Omega$ be the $p\times p$ matrix whose columns are the
  normalized eigenvectors of $\Sigma$, and $\Delta$ be a diagonal
  matrix whose diagonal entries are the eigenvalues of $\Sigma$. Let
  $Q=\Delta^\frac{1}{2}\Omega$.
 Then,  $\vec{Y}=Q\vec{X}+\vec{\mu} \sim N(\vec{\mu},\Sigma)$
\end{enumerate}

This algorithm is used in the software R for simulating multivariate
Normal data. In general, most standard software for statistical
computing (e.g., R, Matlab) have a function for simulating
multivariate Normal data.

\section{Generating Multivariate Poisson Data}\label{sec:poisson}

The $p$-dimensional Poisson distributions is characterized by a mean
(or rate) vector $\vec{\lambda}$ and a covariance matrix
$\Sigma_{Pois}$ that has diagonal elements equal to
$\vec{\Lambda}$.

Sampling from multivariate Poisson distribution has been addressed
massively in the literature, with a major focus on the bivariate
case. It is customary to use the term ``multivariate Poisson" for
any extension of the univariate Poisson distribution where the
resulting marginals are of univariate Poisson form. In other words,
the same term is used to describe different multivariate
distributions, which have in common the property that their
marginals are univariate Poisson.

One of the earliest simulation methods was proposed by
\cite{Krummenauer98,Krummenauer99}. The algorithm first generates
and then convolves independent univariate Poisson variates with
appropriate rates. The author presented a recursive formula to carry
out the convolution in polynomial time. This method enables the
simulation of multivariate Poisson data with \emph{arbitrary}
covariance structure. The main limitations of this method is its
high complexity (the recursions become very inefficient when p
increases). Also, the method does not support negative correlation.

\cite{Minhajuddin04} presented a method for simulating multivariate
Poisson data based on the Negative Binomial - Gamma mixture
distribution. First, a value $k$ is generated from a Negative
Binomial distribution with rate $r$ and
$\Pi=\frac{\lambda}{\lambda+\theta}$. Then, conditional on $k$, a
set of $p$ independent Gamma variates are generated ($X_1,\ldots,
X_p$). The sum of the joint distribution of $k$ and $X_1,\ldots,
X_p$ has a Gamma marginal distribution with rates $r$ and $\lambda$.
The correlation between a pair $X_i$ and $X_j$ ($i\neq j$) is
$\frac{\theta}{\lambda+\theta}$. There are two main drawbacks to
this approach: First, it requires the correlation between each pair
of variates to be identical ($\rho_{ij}=\rho$ for all $i\neq j$).
Second, it does not support a negative correlation.

\cite{Karlis03} points out that the main obstacle limiting the use
of multivariate simulation methods for Poisson data, including the
above-mentioned methods, is the complexity of calculating the joint
probability function. He mentions that the required summations might
be computationally exhausting in some cases, especially when the
dimension $p$ is high.

In a recent paper \cite{Shin07} present a fast method of generating
a bivariate Poisson process with negative correlation. Similar to
the approach of \cite{Minhajuddin04}, their methods is based on
generating two dependent random variables from three independent
random variables. The authors propose a computationally fast
modification to the trivariate reduction that enables generating a
bivariate Poisson with a specified negative correlation.

In this paper, we propose an alternative approach for generating a
$p$-dimensional Poisson vector with covariance matrix
$\Sigma_{Pios}$ and rate $\vec{\Lambda}$. The basic idea is first to
generate a $p$-vector from the multivariate Normal distribution and
then to transform it to Poisson using the Gaussian and Poisson
cumulative distribution functions (CDFs). We use the following
notation for these CDFs, respectively:
\begin{eqnarray}
\Phi(x) & = & \int_{-\infty}^{x}\frac{1}{\sqrt{2\sigma^2}}e^{\frac{-u^2}{2}}du \label{eq:Phi} \\
\Xi(x) & = & \sum_{i=0}^x \frac{e^{-\lambda}\lambda^i}{i!}
\label{eq:Xi}
\end{eqnarray}
Our algorithm proceeds as follows:
\begin{enumerate}
  \item Generate a $p$-dimensional Normal vector $\vec{X^N}$ with mean
  $\vec{\mu}=0$, variance $\vec{\sigma}=1$ and a correlation matrix
  $R^N$.
  \item For each value $X^N_i, i \in {1, 2, ... p}$, calculate the Normal
  CDF:
  $$\Phi(X^N_i)$$.
  \item For each $\Phi(X^N_i)$, calculate the Poisson inverse CDF (quantile) with
  rate $\lambda_i$: $$X^{Pois}_i=\Xi^{-1}(\Phi(X^N_i))$$.
\end{enumerate}

The vector $\vec{X^{Pois}}$ is then a $p$-dimensional Poisson vector
with correlation matrix $R^{Pois}$ and rates $\vec{\Lambda}$.

\textbf{Theorem:} The vector $\vec{X^{Pois}}$ is a $p$-dimensional
vector with Poisson marginals.

\textbf{Proof:} From the properties of multivariate Normal
distribution, $\vec{X^{N}}$ is a $p$-dimensional vector with Normal
marginals. Therefore, by transforming the marginals using the Normal
CDF we obtain a p-dimensional vector $\vec{\Phi(X^N)}$ with uniform
marginals.

We then transform the uniform variates to Poisson variates using the
Poisson inverse CDF, and thus obtain p-dimensional Poisson
marginals.

In the next sections, we experimentally show that the resulting
correlation structure $R^{Pois}$ stochastically equals the desired
correlation structure $R^{N}$.

\subsection{Implementation Issues}

The first two steps are easy to implement (numerical procedures for
computing the Normal CDF are computationally efficient and widely
available). The difficulty arises in the last step, in computing the
Poisson inverse CDF, because of the factorial in the denominator of
$\Xi$ (see equation (\ref{eq:Xi})). Solutions are as follows: For
small values of $\lambda$ (e.g., $\lambda<10$) Poisson variates can
be efficiently simulated by generating exponential (inter-arrival)
random variates and then summing them (see, e.g., \cite{Devroye86}).
For larger values of $\lambda$, ~\cite{Ahrens82} proposed an
algorithm for generating univariate Poisson variates by modifying
Normal variates. An alternative is to approximate the factorial in
equation (\ref{eq:Xi}) using the Sterling approximation. With
today's computational power, this is a feasible possibility.

\section{Graphical Illustration}\label{sec:results}

\begin{sloppypar}
We have not proved that the operations in steps (2)-(3) of our
method maintain the original correlation structure (i.e., that
$\Xi^{-1}(\Phi(\cdot))$ maintains $(R^N)$. We next show, in a series
of experiments, that this indeed is the case. In particular, we show
that the multivariate Poisson data that are generated by our method
have Poisson marginal distributions and a correlation structure
equal to the one specified in step (1).
\end{sloppypar}

We use the software R to generate the data, with the functions
\emph{\texttt{mvrnorm}} to generate multivariate Normal vectors,
\emph{\texttt{pnorm}} for the Normal CDF and \emph{\texttt{qpois}}
for the Poisson inverse-CDF. The code in given in the Appendix.

For each of the cases in sections 4.1-4.2 below we generate 50,000
3-dimensional Poisson variates with rate vectors
$\vec{\Lambda}=\{\lambda_1, \lambda_2, \lambda_3\}$ and correlation
matrix $R$. When the variance is different from 1, we provide the
covariance matrix $\Sigma$ which has elements: $\forall i,j \in p:
Cov(X_i,X_j)=\rho_{ij}\sigma_i\sigma_j$, where $\rho_{ij}$ is the
correlation between $x_i$ and $x_j$.

\subsection{Constant rate and equal correlations}

We start with the simple case of a multivariate Poisson distribution
with equal pairwise correlations ($\rho=0.4$) and a constant rate
vector ($\lambda_1=\lambda_2=\lambda_3=2$). In other words, we
generate 3-dimensional Poisson data with the following correlation
and covariance matrices:

\begin{eqnarray}\label{eq:const}
\begin{tabular}{p{5cm}}
  \[\Sigma= \left(
   \begin{array}{ccc}
     2 & 0.8 & 0.8 \\
     0.8 & 2 & 0.8 \\
     0.8 & 0.8 & 2 \\
   \end{array}
 \right)\]
 \\
\end{tabular}\begin{tabular}{p{5cm}}
  \[R= \left(
   \begin{array}{ccc}
     1 & 0.4 & 0.4 \\
     0.4 & 1 & 0.4 \\
     0.4 & 0.4 & 1 \\
   \end{array}
 \right)\]
\end{tabular}
\end{eqnarray}

\begin{figure}[h]
\centering
\includegraphics[width=6cm]{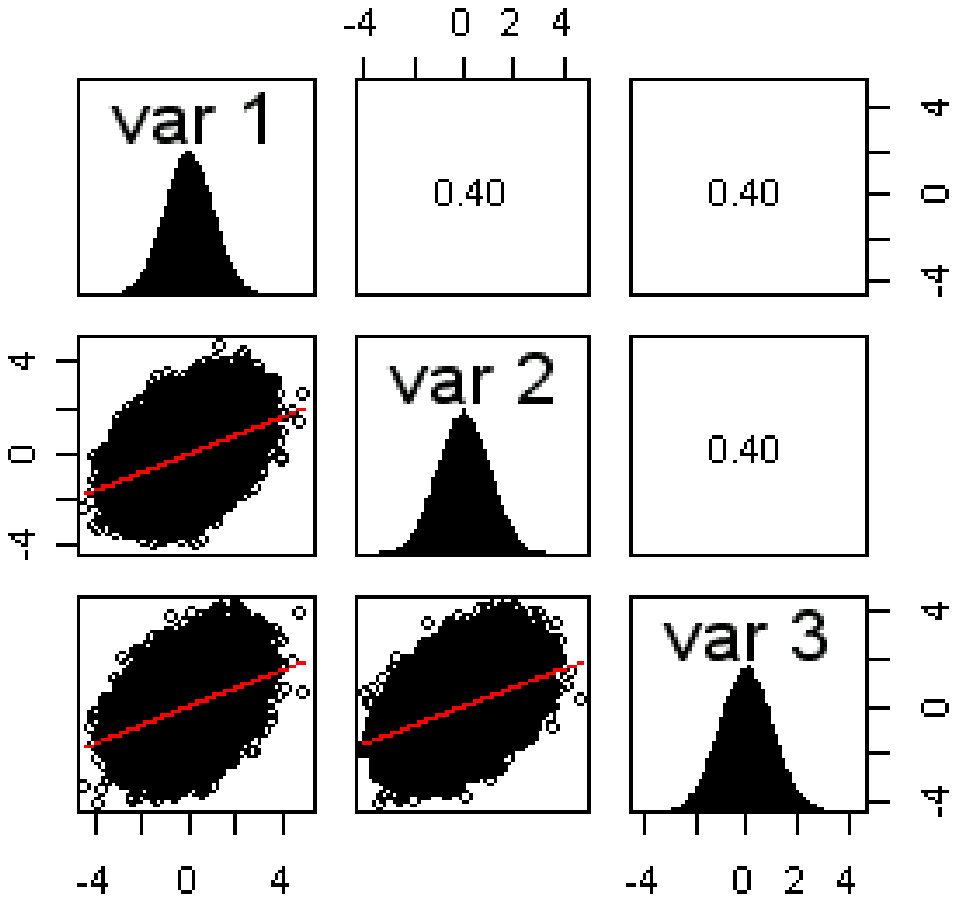}
\includegraphics[width=6cm]{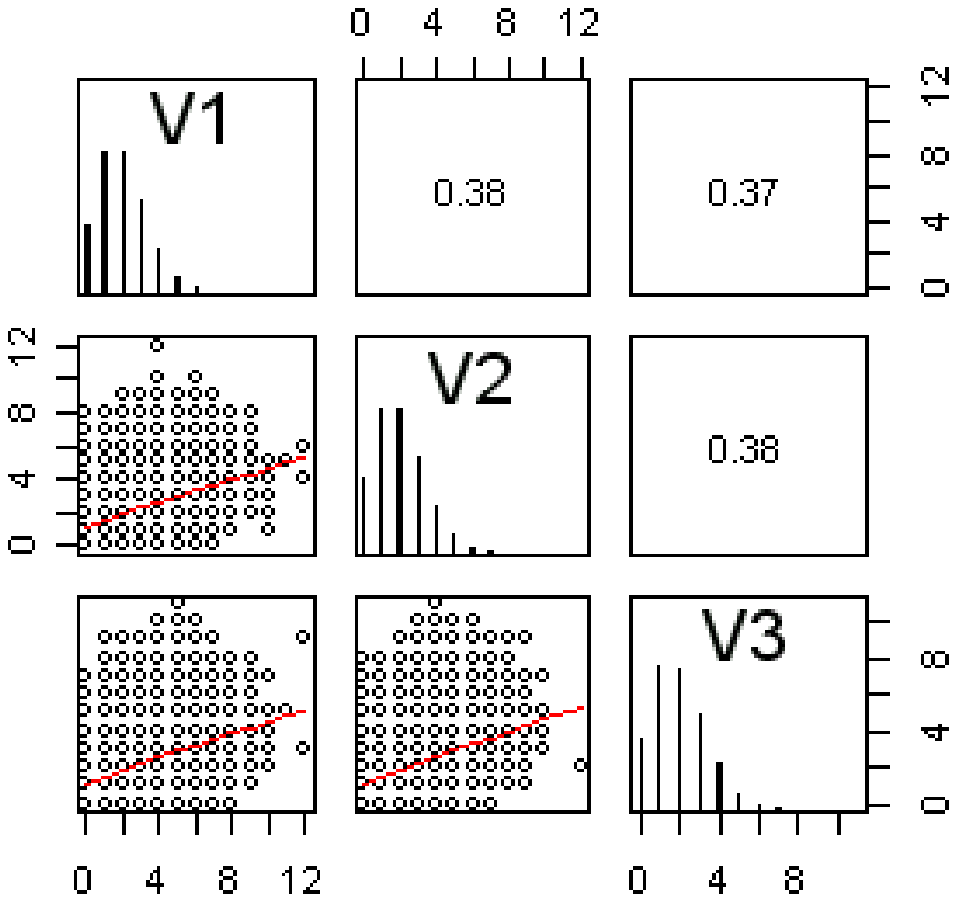}
\caption{Comparing the simulated Normal (left) and Poisson (right)
multivariate data with correlation matrix given in equation
\ref{eq:const}} \label{fig:Test1}
\end{figure}

\begin{figure}[h]
\centering
\includegraphics[width=5cm]{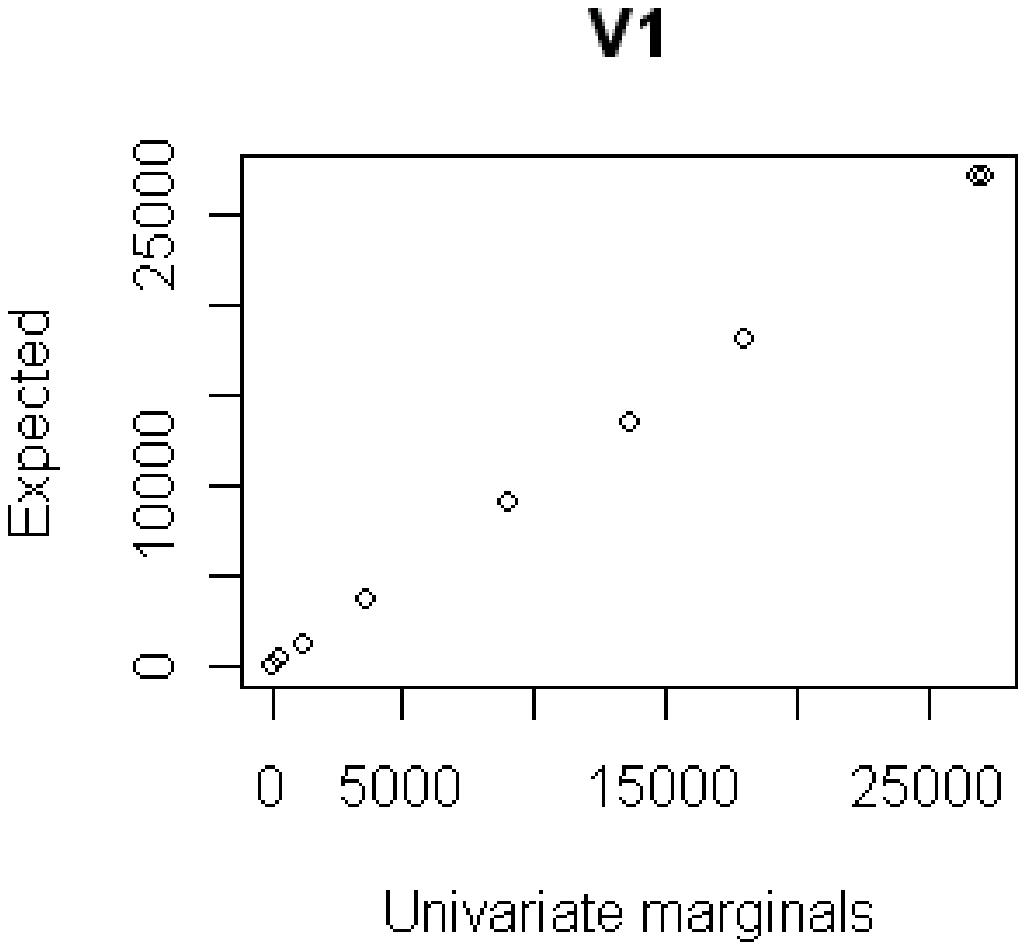}
\includegraphics[width=5cm]{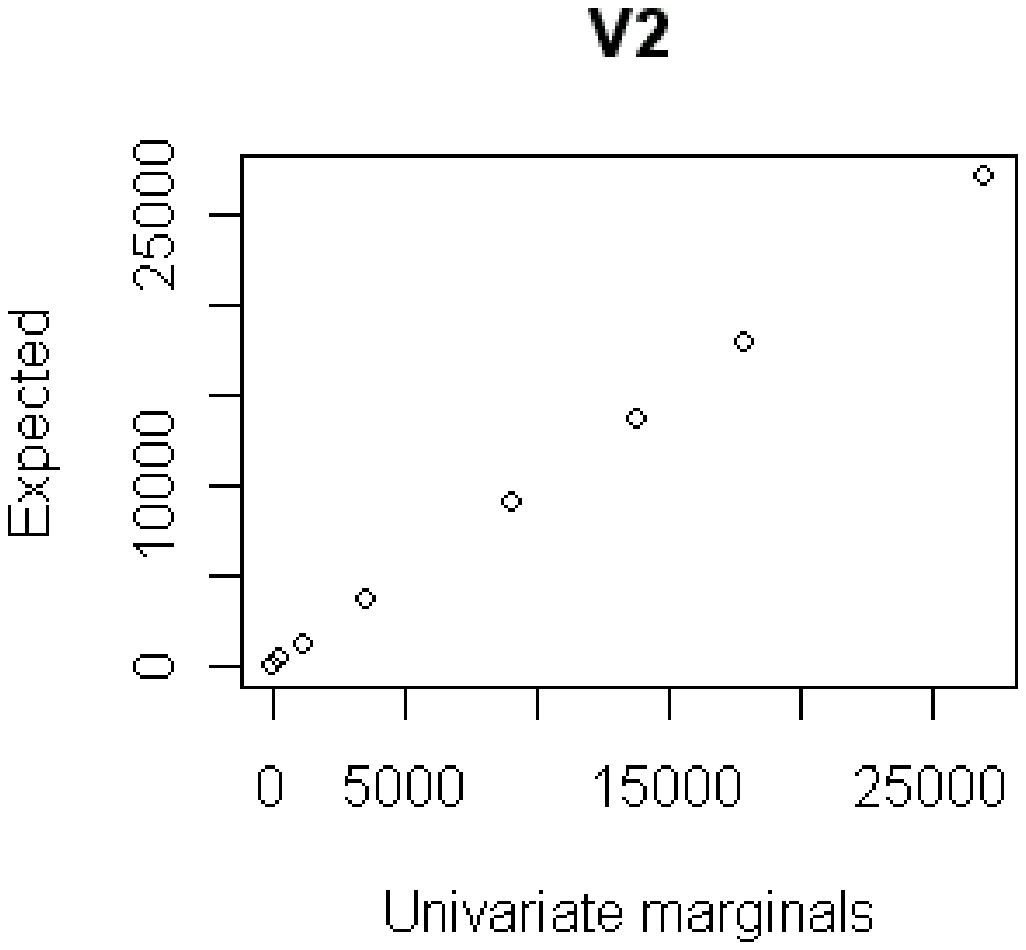}
\includegraphics[width=5cm]{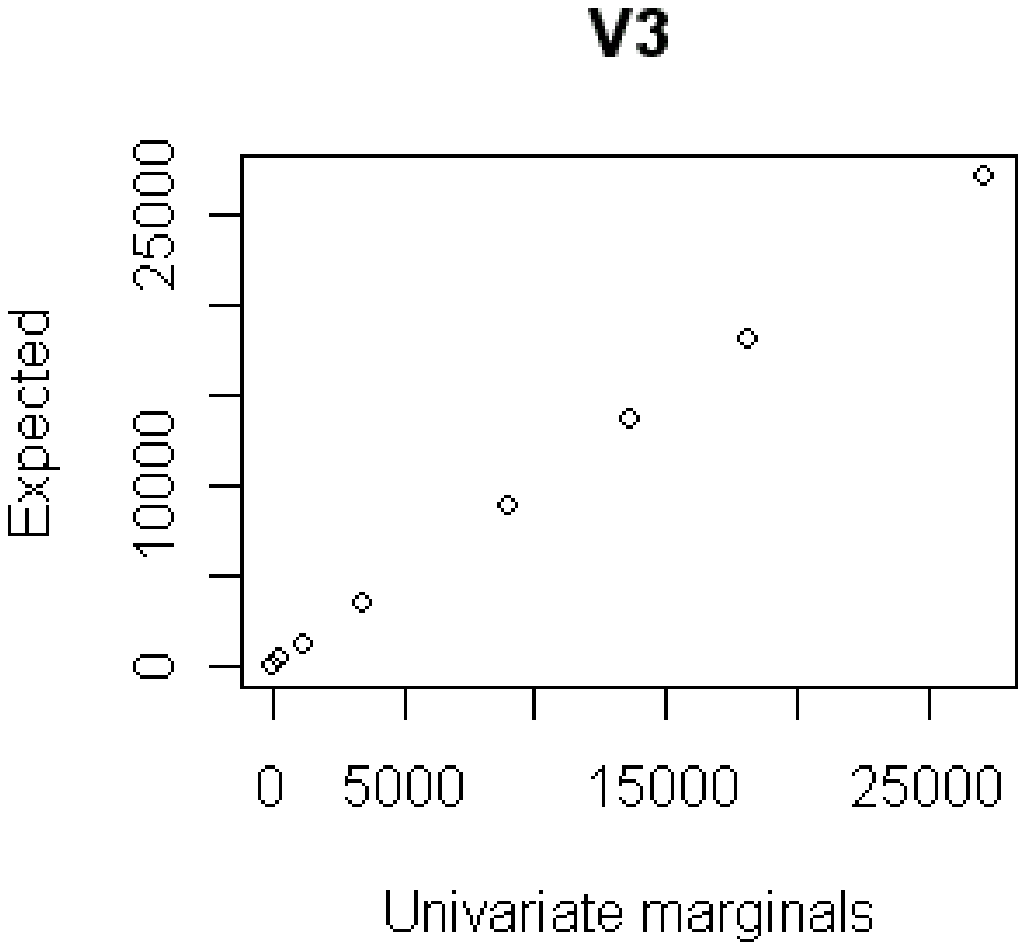}
\caption{Comparing the marginal counts to Poisson expected values}
\label{fig:qqplot1}
\end{figure}

Figure \ref{fig:Test1} compares the initial multivariate Normal data
(left panel) and the resulting multivariate Poisson data (right
panel). Each panel contains scatter plots for each of the three
pairs of series (bottom-left), the estimated correlation
coefficients (upper-right), and histograms of the marginal
distributions (diagonal). We see that the Normal and Poisson data
exhibit Normal and Poisson marginals, respectively, and that both
datasets share the same correlation structure, as desired. To
further examine the fit of each of the three marginals to a Poisson
distribution we provide scatter plots of the marginal counts
compared to their expected values under a Poisson distribution in
Figure \ref{fig:qqplot1}.

\subsection{Non-constant rate, non-constant correlations, and negative correlations}

We next examine the case of a non-constant Poisson rate vector,
unequal pairwise correlations, and negative correlations. All tests
applied to the simulated data indicate that the pre-specified
correlation matrix is maintained and that the marginal distributions
are each Poisson. For sake of brevity we present only results for
one configuration, but the same results were obtained for a wide
range of vector rates and correlation matrices. In this test we set
$\vec{\Lambda}=\{5, 10,
15\}$ and \\

\begin{eqnarray}\label{eq:non_const}
\begin{tabular}{p{5cm}}
  \[\Sigma= \left(
   \begin{array}{ccc}
     5 & -2.83 & 3.33 \\
     -2.83 & 10 & 4.9 \\
     3.33 & 4.9 & 15 \\
   \end{array}
 \right)\]
\end{tabular}\begin{tabular}{p{5cm}}
  \[R= \left(
   \begin{array}{ccc}
     1 & -0.4 & 0.4 \\
     -0.4 & 1 & 0.5 \\
     0.4 & 0.5 & 1 \\
   \end{array}
 \right)\]
\end{tabular}
\end{eqnarray}

\begin{figure}[h]
\centering
\includegraphics[width=6cm]{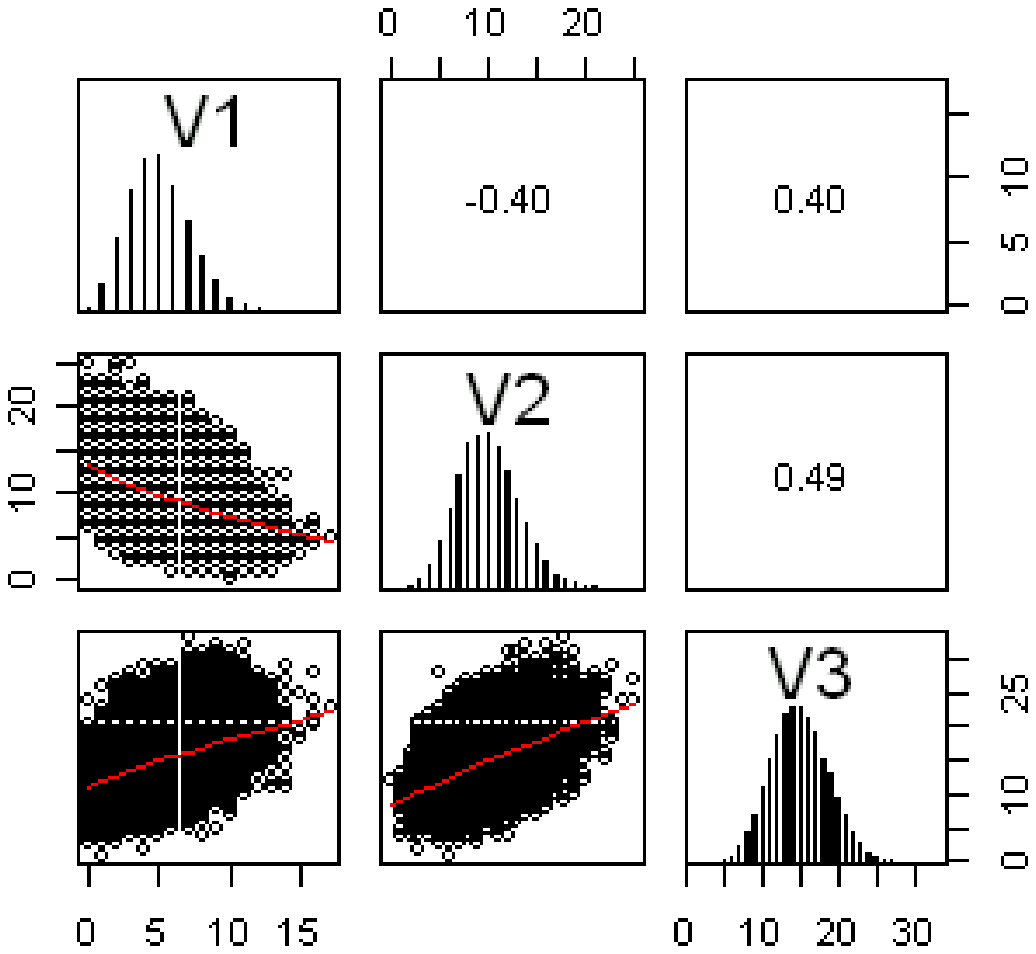}
\caption{Test 2: Simulating multivariate Poisson data with
correlation matrix given in equation \ref{eq:non_const}}
\label{fig:Test4Pois}
\end{figure}

\begin{figure}[h]
\centering
\includegraphics[width=5cm]{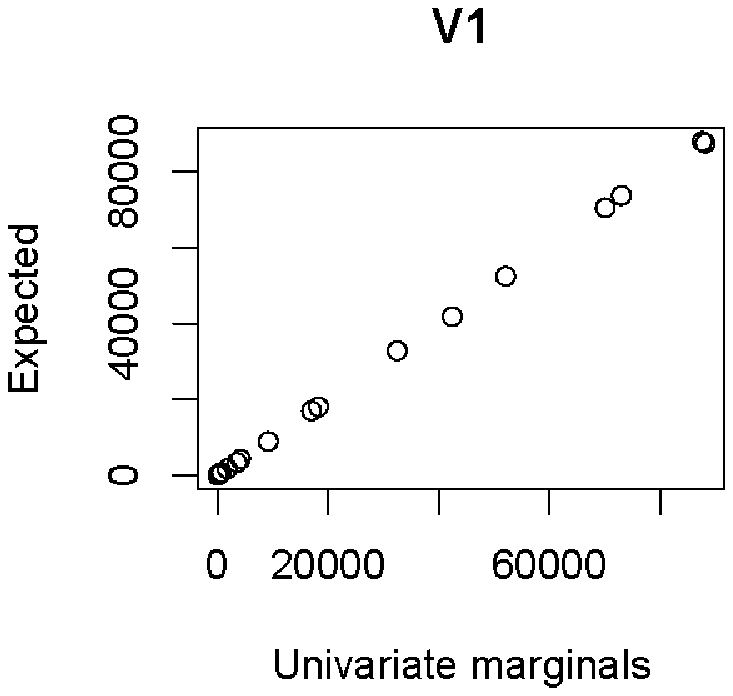}
\includegraphics[width=5cm]{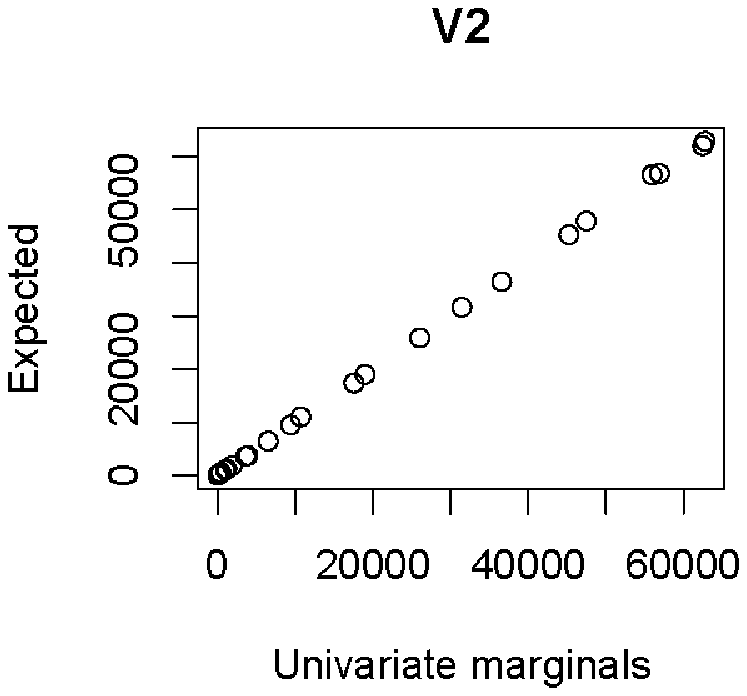}
\includegraphics[width=5cm]{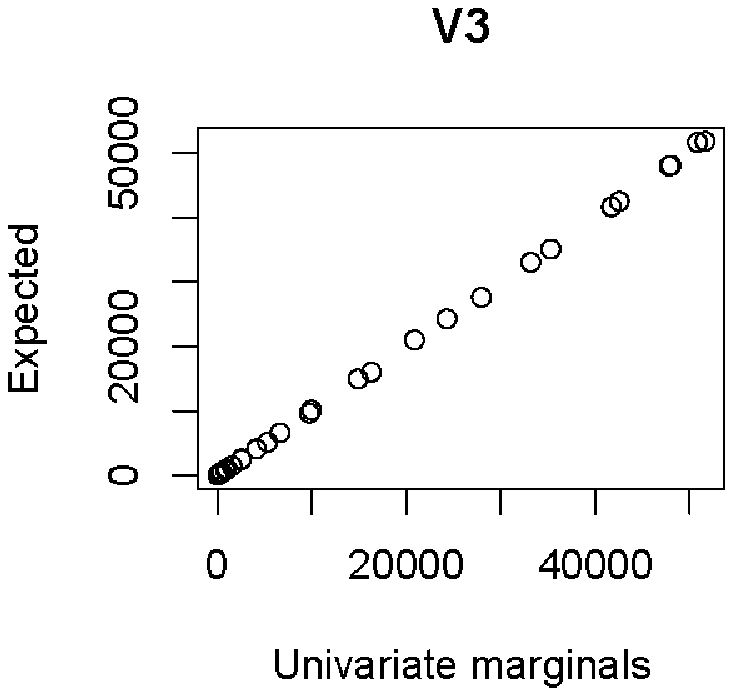}
\caption{Comparing the Poisson marginals to the expected values}
\label{fig:qqplot2}
\end{figure}

The simulated 3-variate Poisson data are shown in Figure
\ref{fig:Test4Pois} and the goodness of fit to a Poisson
distribution is shown in the observed vs. expected scatter plots in
Figure \ref{fig:qqplot2}. We see that the estimated correlations are
very close to those in the pre-specified $R$ (including the negative
correlations) and that each of the marginal distributions closely
fits a Poisson distribution.

\subsection{Multivariate Poisson with Low Rates}

\begin{figure}[h]
\centering
\includegraphics[width=8cm]{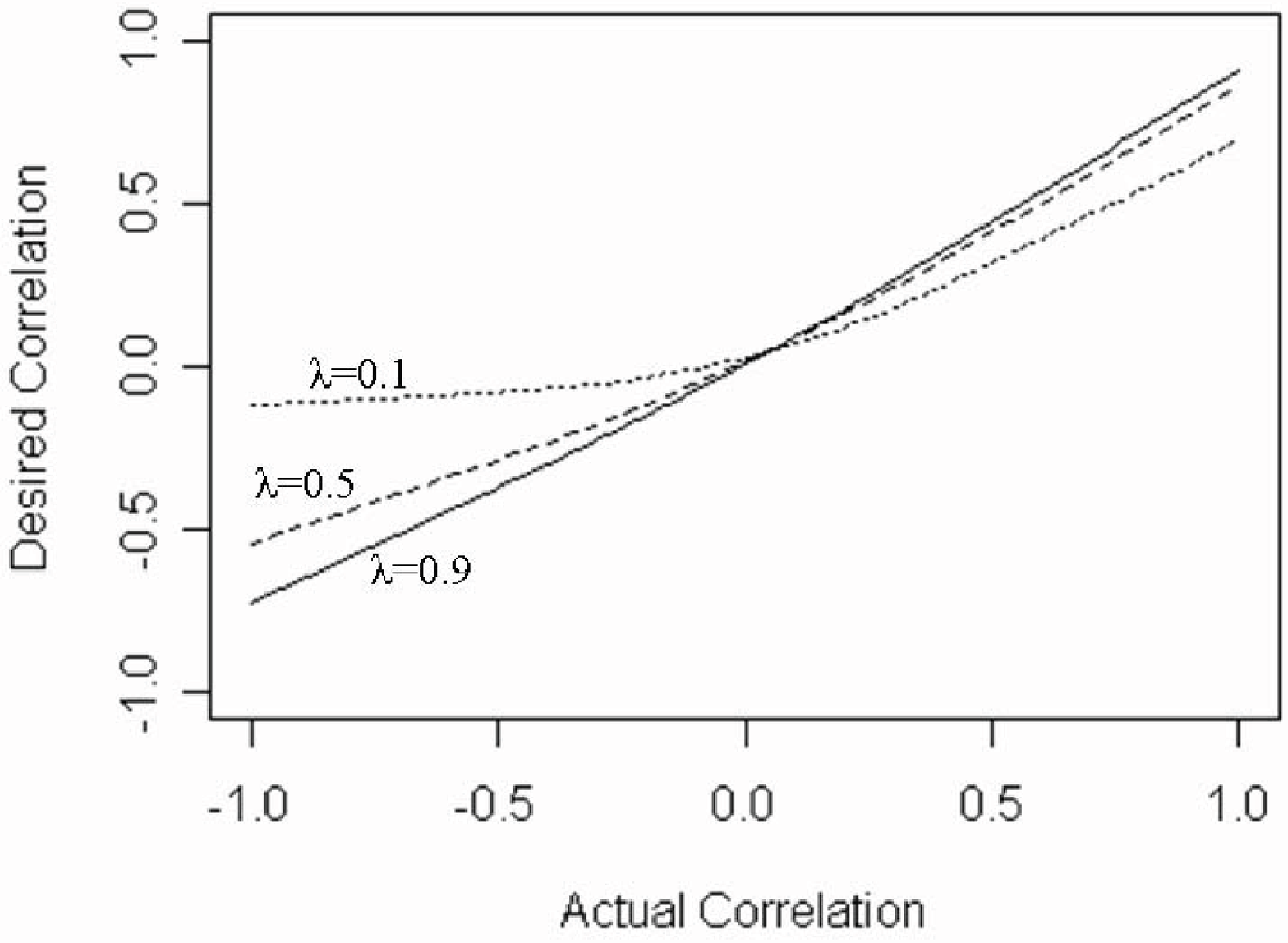}
\caption{Comparing the desired correlation to the resulting actual
correlation for low rates} \label{fig:des_act}
\end{figure}

Finally, we examine our approach for low rates ($\lambda\leq4$). It
is well known that the Poisson distribution with high rates
($\lambda>5$) is approximately a Normal distribution, and we thus
expect the actual correlation of the multivariate Poisson random
variables to match the desired correlation. However, as shown in
\cite{whitt1976bdg}, the feasible correlation between two random
Poisson variables is no longer in the range [$-1,1$], but rather
[$min\_corr=corr(\Xi_{\lambda_1}^{-1}(U),\Xi_{\lambda_2}^{-1}(1-U)),
max\_corr=corr(\Xi_{\lambda_1}^{-1}(U),\Xi_{\lambda_2}^{-1}(U))$].
In fact, \cite{Shin07} show that when $\lambda_1, \lambda_2
\rightarrow 0$ the minimum feasible correlation $min\_corr
\rightarrow 0$. Therefore, our transformation maps a correlation
range of [$-1,1$] (multivariate normal) to a much smaller range
[$min\_corr \geq -1, max\_corr \leq 1$]. Figure \ref{fig:des_act}
illustrates the relationship between the desired correlation and the
resulting actual correlation when generating bivariate Poisson data
with low rates ($\lambda<1$).

We find that the relationship between the desired correlation and
the actual correlation can be approximated by an exponential form:
\[R^{Pois}=a\times e^{bR^{N}}+c,\] where the coefficient $a$, $b$ and
$c$ can be estimated from the points ($\min\_corr$, -1),
($\max\_corr$, 1) and (0, 0):

\begin{eqnarray}
  \nonumber && a=-\frac{max\_corr\times min\_corr}{max\_corr+min\_corr} \\
  \nonumber && b=log (\frac{max\_corr+a}{a}) \\
  \nonumber && c=-a
\end{eqnarray}

\begin{figure}[h]
\centering
\includegraphics[width=8cm]{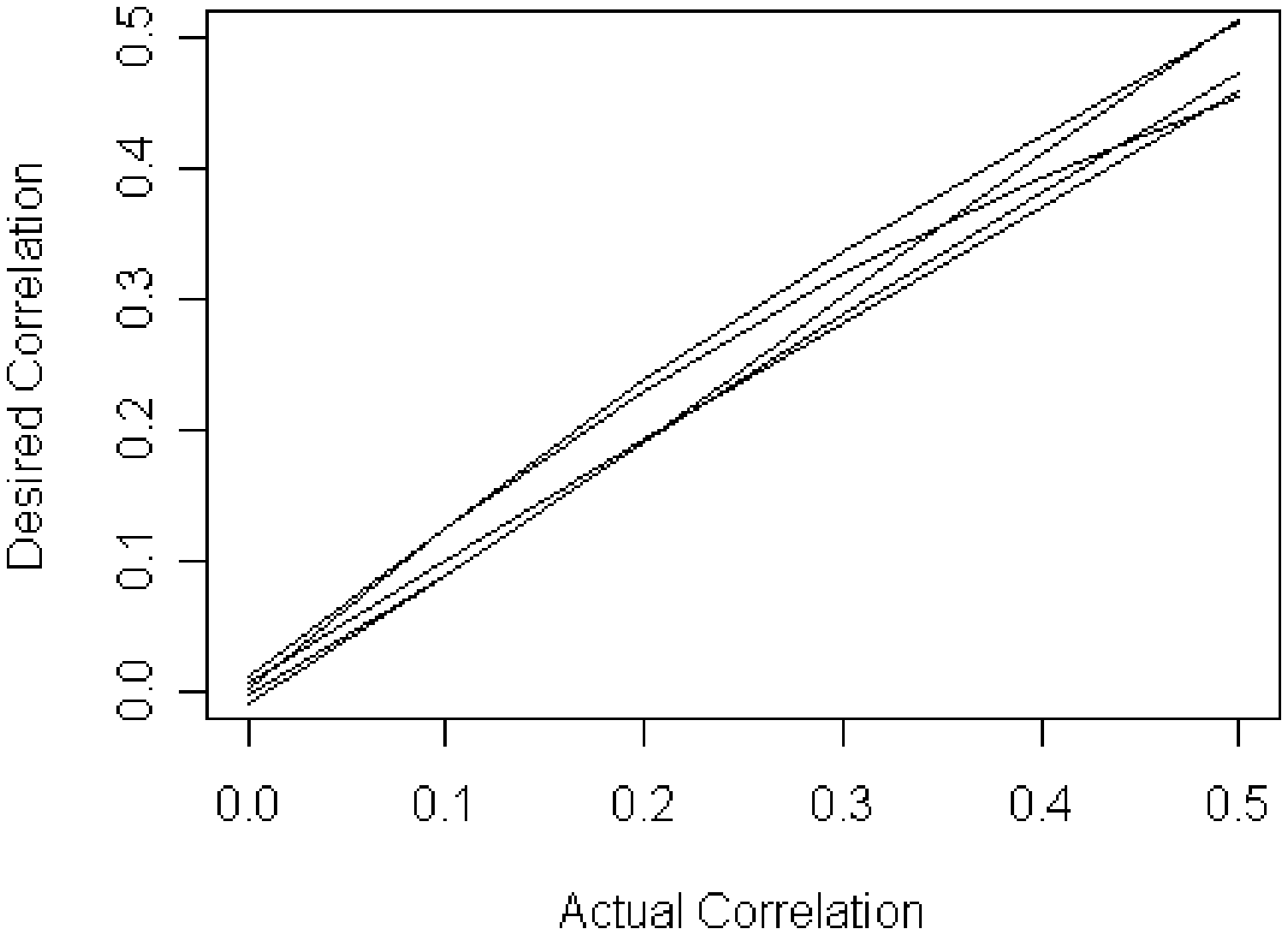}
\caption{Comparing the desired correlation to the corrected actual
correlation} \label{fig:correct}
\end{figure}

\begin{sloppypar}
Figure \ref{fig:correct} illustrates the simulation performance when
using the above approximation to correct for the distortion in the
resulting correlation. This is illustrated for the bivariate Poisson
case with rates that range $(\lambda_1, \lambda_2) \in {(0.1, 0.1),
(0.1, 0.5), (0.5, 0.5), (0.5, 0.9), (0.9, 0.9)}$. The absolute
difference between the actual and desired correlation is less then
0.05.
\end{sloppypar}

\section{Conclusions}\label{sec:conclusion}

Simulating multivariate Poisson data is essential in many real-world
applications in a wide range of fields such as healthcare,
marketing, management science, and many others. Current simulation
methods suffer from computational limitations and restrictions on
the correlation structure, and therefore are rarely used.

In this paper we propose an elegant way to generate multivariate
Poisson data based on a multivariate Normal distribution with a
pre-specified correlation matrix and Poisson rate vector. Because
multivariate Normal and univariate Poisson simulators are
implemented in many standard statistical software packages,
implementing our method requires only a few lines of code.

We show that our method works well for different correlation
structures (both negative and positive; and varying values) and for
non-constant rates. We show that the method is highly accurate in
terms of producing Poisson marginal distributions and the
pre-specified correlation matrix.

\section*{Acknowledgement}
We thank Professor Michael Fu from the Smith School of Business,
University of Maryland College Park for his useful comments.

The work was partially supported by NIH grant RFA-PH-05-126.

\newpage
\bibliographystyle{chicago}

\begin{thebibliography}{}

\bibitem[\protect\citeauthoryear{Ahrens and Dieter}{Ahrens and
  Dieter}{1982}]{Ahrens82}
Ahrens, J.~H. and U.~Dieter (1982).
\newblock Computer generation of poisson deviates from modified normal
  distributions.
\newblock {\em ACM Transactions on Mathematical Software\/}~{\em 8}, 163–179.

\bibitem[\protect\citeauthoryear{Devroye}{Devroye}{1986}]{Devroye86}
Devroye, L. (1986).
\newblock {\em Non-Uniform random variate generation}, Chapter The Poisson
  Proess, pp.\  245--250.
\newblock Springer-Verlag.

\bibitem[\protect\citeauthoryear{Hernádvölgyi}{Hernádvölgyi}{1998}]{lgyi98}
Hernádvölgyi, I.~T. (1998).
\newblock Generating random vectors from the multivariate normal distribution.
\newblock Working paper. Available at citeseer.ist.psu.edu/78434.html.

\bibitem[\protect\citeauthoryear{Karlis}{Karlis}{2003}]{Karlis03}
Karlis, D. (2003).
\newblock An em algorithm for multivariate poisson distribution and related
  models.
\newblock {\em Journal of Applied Statistics\/}~{\em 30\/}(1), 63--77.

\bibitem[\protect\citeauthoryear{Krummenauer}{Krummenauer}{1998}]{Krummenauer9%
8} Krummenauer, F. (1998).
\newblock Limit theorems for multivariate discrete distributions.
\newblock {\em Metrika\/}~{\em 47\/}(1), 47--69.

\bibitem[\protect\citeauthoryear{Krummenauer}{Krummenauer}{1999}]{Krummenauer9%
9} Krummenauer, F. (1999).
\newblock Efficient simulation of multivariate binomial and poisson
  distributions.
\newblock {\em Biometrical Journal\/}~{\em 40\/}(7), 823--832.

\bibitem[\protect\citeauthoryear{Minhajuddin, Harris, and Schucany}{Minhajuddin
  et~al.}{2004}]{Minhajuddin04}
Minhajuddin, A.~T., I.~R. Harris, and W.~Schucany (2004).
\newblock Simulating multivariate distributions with specific correlations.
\newblock {\em Journal of Statistical Computation \& Simulation\/}~{\em
  74\/}(81), 599--607.

\bibitem[\protect\citeauthoryear{Ripley}{Ripley}{1987}]{Ripley87}
Ripley, B.~D. (1987).
\newblock {\em Stochastic Simulation}, Chapter Multivariate Distibutions, pp.\
  98--100.
\newblock John Wiley \& Sons.


\bibitem[\protect\citeauthoryear{Shin, Pasupathy}{Shin and Pasupathy}{2007}]{Shin07}
Shin k., and Pasupathy R. (2007).
\newblock {\em A Method For Fast Generation of Bivariate Poisson Random Vectors},
Procceeding of the 2007 Winter Simulation Conference, pp.\
  472--479.

\bibitem[\protect\citeauthoryear{Whitt, W}{Whitt}{1976}]{whitt1976bdg}
Whitt, W.. (2007).
\newblock {\em Bivariate Distributions with Given Marginals},
The Annals of Statistics 4(6), pp.\
  1280--1289.
\end{thebibliography}

\newpage
\begin{center}
    {\bf APPENDIX: Generating Multivariate Poisson Data in R}
\end{center}

\begin{verbatim}
    # Generate a p-dimensional Poisson
    # p         = the dimension of the distribution
    # samples   = the number of observations
    # R         = correlation matrix p X p
    # lambda    = rate vector p X 1
GenerateMultivariatePoisson<-function(p, samples, R, lambda){
        normal_mu=rep(0, p)
        normal = mvrnorm(samples, normal_mu, R)
        pois = normal
        p=pnorm(normal)
        for (s in 1:p){pois[,s]=qpois(p[,s], lambda[s])}
        return(pois)
}
\end{verbatim}

\begin{verbatim}
    # Correct initial correlation between a
    # certain pair of series
    # lambda1   = rate of first series
    # lambda2   = rate of second series
    # r         = desired correlation
CorrectInitialCorrel<-function(lambda1, lambda2, r){
    samples=500
    u = runif(samples, 0, 1)
    lambda=c(lambda1,lambda2)
    maxcor=cor(qpois(u, lambda1), qpois(u, lambda2))
    mincor=cor(qpois(u, lambda1), qpois(1-u, lambda2))
    a=-maxcor*mincor/(maxcor+mincor)
    b=log((maxcor+a)/a, exp(1))
    c=-a
    corrected=log((r+a)/a, exp(1))/b
    corrected=ifelse ((corrected>1 | corrected<(-1)),
                       NA, corrected)
    return(corrected)
}
\end{verbatim}

\end{document}